\documentclass{edp-jp4}
\usepackage{graphicx}
\usepackage{amsmath,amssymb}
\usepackage{indentfirst}
\newcommand{\dip}[1]{\times 10^{#1}}

\newcommand{\ket}[1]{\mbox{$\left|#1\right\rangle$}}
\newcommand{\bvec}[1]{{\mathbf{#1}}} 
\newcommand{\vecx}{{\mathbf{e}}_x}
\newcommand{\vecy}{{\mathbf{e}}_y}
\newcommand{\vecz}{{\mathbf{e}}_z}

\newcommand{\muo}{\mu_0}
\newcommand{\Bot}{B_{0\perp}}  
\newcommand{\Bol}{B_{0\parallel}}  
\newcommand{\Bw}{B_{\text{w}}} 
\newcommand{\Btot}{B_{\text{t}}} 
\newcommand{\Bs}{B_{\text{s}}} 
\newcommand{\ro}{\rho_0} 

\begin{document}

\title{Using magnetic chip traps to study Tonks-Girardeau quantum gases}
\author{Jakob Reichel} \address{Max-Planck-Institut f\"ur Quantenoptik
  and Sektion Physik der Ludwig-Maximilians-Universit\"at,
  Schellingstr. 4, D-80799 M\"unchen, Germany} \author{Joseph H.
  Thywissen} \address{McLennan Physical Laboratories, University of
  Toronto, 60 Saint George Street, Toronto, Ontario, M5S 1A7 Canada}
\maketitle
\begin{abstract}
  We discuss the use of microfabricated magnetic traps, or ``chip traps,'' to study
  \mbox{(quasi-)}one-dimensional quantum gases.  In particular, we discuss the
  feasibility of studying the Tonks-Girardeau limit, in which
  the gas is strongly interacting.  We review the scaling of the
  oscillation frequencies of a chip trap, and show that it seems feasible to attain
  a Tonks-Girardeau parameter as large as 200. The primary difficulty
  of this approach is detection, since the strongly interacting limit
  occurs for low densities. We propose a way to ``freeze'' the
  distribution, and then measure it with a single-atom detector. This method
  can also be applied to optical dipole traps.
\end{abstract}

\section{Introduction}
\label{sec:intro}

The most exciting regime of 1D neutral gases is the Tonks-Girardeau (TG) regime\cite{TGearly,Olshanii98,Petrov00,Dunjko,Girardeau01,MoreTonksTheory,Shlyapnikov03}, wherein the gas is {\em strongly interacting}. Three-dimensional strongly interacting gases require $n a^3 \gtrsim 1$, where $n$ is the density
of the gas, and $a$ is the s-wave scattering length. However, in this regime the three-body collisional loss of trapped neutral atoms can be prohibitively high.  By contrast, 1D gases are strongly interacting in a {\em low-}density regime, which is perhaps more experimentally accessible.  

At the time that this manuscript was originally written, although much theoretical work had been done on the TG limit, this highly correlated (\textit{i.e.}, beyond mean field) regime had never been realized -- with neutral atoms or with any other constituent,  Since then, two reports have been made \cite{PhillipsTonks,BlochTonks}. Both of these reports concern a Bose gas trapped in a two-dimensional optical array of one-dimensional traps. The data in \cite{PhillipsTonks} presents a reduction in the three-body loss rate, as predicted by Shlyapnikov {\it et al.} \cite{Shlyapnikov03} to indicate a reduction in the correlation function $g^{(3)}$ at short range. The data in \cite{BlochTonks} measure the time-of-flight distribution after release from the trap, and compare to theoretical expectations.

In this article, we propose a new method -- based on the combination of a strong, anisotropic trap, and a single-atom detector -- by which one can create an atomic ensemble in the TG regime and study its position distribution and correlations. In such a study, the crossover from ideal bosonic to ideal fermionic behavior -- which is the most striking aspect of the TG quantum gas -- would be immediately apparent.  We start by analyzing, in \S\ref{sec:scaling}, the scaling properties of magnetic microchip traps to determine the maximum confinement that can be achieved. In \S\ref{sec:TG}, we discuss a specific trap layout for the creation of a TG gas. The detection of such a gas is considered in \S\ref{sec:detection}.  Finally, in \S\ref{sec:conclusion}, we discuss the practical issues with such an experiment, and conclude.

\section{Scaling properties of chip-based magnetic traps}
\label{sec:scaling}

\subsection{Trapping fields from planar current distributions}
When a magnetic potential is created by a system of wires with characteristic size $s$ and carrying a current $I$, the trapping field gradient and curvature scale respectively as $I/s^2$ and $I/s^3$ when $s$ is decreased \cite{Weinstein95}.  Therefore, traps that replace the customary field coils by thin wires on substrates can provide  more strongly confining potentials with much less power dissipation than ``traditional'' traps using macroscopic coils. This is the basic idea of microchip traps, also known as ``atom chips''. The properties of such traps have been reviewed recently \cite{FolmanReview02,Reichel02}.  This section focuses on elongated traps with strong transverse confinement, i.e., high transverse oscillation frequency. We first recall how such traps can be constructed with wires and homogeneous external fields, and then discuss the strongest confinement that can be realistically expected for such a trap.

\subsubsection{Thin wires and two-dimensional confinement}
\label{sec:thinWires}

In a chip trap, all field gradients are produced by wires. Consider an infinitely thin wire along the $z$ axis, carrying a current $I$. This wire creates a magnetic field $\bvec{\Bw}$, which has the gradient
\begin{equation}
  \Bw'(\rho) =-\frac{\muo}{2\pi}\frac{I}{\rho^2}\label{eq:BSgrad}\\
\end{equation}
at a distance $\rho$ (in cylindrical coordinates). The wire field alone does not provide trapping because it does not possess a minimum. One way to construct a trapping potential from this field is to add a uniform external field $\bvec{\Bot}$ perpendicular to the wire axis $\vecz$. The sum of the two fields, $\bvec{\Btot}=\bvec{\Bw}+\bvec{\Bot}$, is zero on a straight line parallel to the $z$ axis at a distance $\ro$ from the wire axis:
\label{sec:trapPrinc}
\begin{displaymath}
  \ro    = \frac{\muo}{2\pi}\frac{I}{\Bot}\,;
\end{displaymath}
this line forms the axis of a two-dimensional trap. Near this trap axis, the field modulus grows linearly and its gradient is $\Bw'(\ro)$, \textit{i.e.} equal in magnitude to that of the wire alone:
\begin{equation}
  \label{eq:gradCenter}
  |\Btot'(\ro)| = \frac{2\pi}{\muo}\frac{\Bot^2}{I}\,.
\end{equation}

Thus, the superposition of the wire and external fields create a two-dimensional quadrupole trap, or atom guide, with a transverse restoring force proportional to $\Btot'(\ro)$. Arrangements of several parallel wires, either with or without external fields, can also be used to create such guides, as discussed in \cite{Thywissen99}.

\subsubsection{Finite wire width}
\label{sec:broad}

The finite cross-section of a real wire limits the field gradient $\Bw'$ that can be reached for a given current. In the case of a wire with circular cross-section of diameter $w$, the field outside the conductor is identical to that of an infinitely thin wire centered on the cylinder axis; as the trap must be placed outside the conductor, the maximum gradient is
\begin{equation}
|\Bs'|=\frac{2\muo}{\pi}\,\frac{I}{w^2}\,.
\label{eq:gradMax}
\end{equation}

For a wire of rectangular cross-section with zero height, but nonzero width $w$, the field gradient at the wire surface has exactly the same value \cite{Reichel02}. Analytical formulas also exist for the field and gradient as a function of distance $x$ from the surface of such a wire:
\begin{equation}
  B(x)=\frac{\mu_0}{\pi}\frac{I}{w}\,\textrm{arccot}\frac{w}{2x}=
  \frac{\mu_0}{\pi}\frac{I}{w}\left(\frac{\pi}{2}-\arctan\frac{2x}{w}\right)\,,
  \label{eq:BBroad}
\end{equation}
\begin{equation}
  B'(x)=-\frac{\mu_0}{2\pi}\frac{I}{x^2+(w/2)^2}\,.
  \label{eq:gradBBroad}
\end{equation}

\subsubsection{Finite current density}

In Eq.~\eqref{eq:gradMax}, $\Bs'$ is proportional to the current density $j$ in the wire. Indeed, with $j=\beta I/w^2$ ($\beta=1$ for rectangular cross-section and $\beta=4/\pi$ for circular cross-section), we have $|\Bs'|=2\muo/\pi\times j/\beta$. Therefore, it is necessary to work at the highest possible current density if strong confinement is required. For thick wires ($w\gtrsim 1\ldots 10~\mu$m) current is limited by the total power of ohmic heating, and reducing the wire cross-section enables higher current densities. However, for thin wires ($w\lesssim 1\ldots 10~\mu$m) the maximum current density no longer increases, but becomes independent of $w$ \cite{Reichel02}. Consequently, the maximum field gradient also saturates. The highest reported current densities lie between $2\times 10^{11}$~A/m$^2$ (at room temperature \cite{Lev03}) and $10^{12}$~A/m$^2$ (with liquid nitrogen cooling \cite{Drndic98}), leading to maximum gradients in the $10^5$~T/m region. As far as strength of confinement is concerned, it is desirable to work with wires just thin enough to achieve these current densities, \textit{i.e.} $w\sim 1\ldots 10~\mu$m.

\subsection{Maximum transverse trap frequency}
\label{sec:maxFreq}
It is well known that storage time in quadrupole traps is limited by spin depolarisation (Majorana transitions). We must therefore extend our discussion to traps with nonzero minima. In the two-dimensional trap discussed above, a nonzero field in the minimum can be obtained by adding a ``guiding field'' $\Bol$ along the trap axis $\vecz$. The dependence of the field modulus near the minimum is now quadratic instead of linear, and the trap has a well-defined transverse oscillation frequency, $\omega$. Taking into account the limited current density discussed above, we now estimate the maximum field curvature and trap frequency.

Near the trap center, the total field is well approximated by
\begin{displaymath}
  B(\hat\rho)=\Bol+\frac{{B'}^2}{2\Bol}\hat\rho^2\,,
\end{displaymath}
where $B'$ is the field gradient in the trap center, and $\hat\rho$ is measured from the axis defining the trap center. The transverse frequency is then given by
\begin{equation}
  \omega_\perp=\sqrt{\frac{\mu_{\rm m}}{m}\frac{{B'}^2}{\Bol}}\,,
\label{eq:omega}
\end{equation}
where $\mu_{\rm m}$ is the magnetic moment and $m$ the mass of the atom.
To avoid Majorana losses, $\Bol$ must be proportional to the trap frequency:
\begin{equation}
  \Bol=\alpha\omega_\perp\,,
\label{eq:IPBias}
\end{equation}
with $\alpha\sim 1.7\dip{-10}$~Ts to obtain a spin flip probability of about $10^{-6}$ per oscillation period \cite{Thywissen99,Gov00}. Eliminating $\Bol$ in \eqref{eq:omega} and assuming a gradient $B'=(\mu_0/4)j$ (\textit{i.e.} half the value at the surface of a circular wire, cf.~\eqref{eq:gradMax}) yields
\begin{equation}
  \omega_\perp=\alpha_j\,j^{\frac{2}{3}}
\label{eq:omegaj}
\end{equation}
with
\begin{displaymath}
\alpha_j=\left (\frac{\mu_{\rm m}}{\alpha m}\right )^{\frac{1}{3}}
         \left (\frac{\mu_0}{4}\right )^{\frac{2}{3}} \,.
\end{displaymath}
For $\mu_{\rm m}=\mu_B$ and $m=1.44\dip{-25}$~kg (mass of $^{87}$Rb),
the numerical value of this constant is $\alpha_j=2\pi \times
5.0\dip{-2}$~m$^{4/3}$s$^{-1}$A$^{-2/3}$. The maximum possible
oscillation frequency is obtained by inserting the maximum current
density into \eqref{eq:omegaj}. With $j=10^{11}$~A/m$^2$, the result
for the $\ket{F=2,m=2}$ state of $^{87}$Rb is $\omega_{\max} = 2\pi
\times 1.1$~MHz, with a corresponding ground state size ($1/e$ radius
of $|\Psi|^2$) of $\delta x=10$~nm, and results from a gradient
$B'=3.1\dip{4}$~T/m. The value of the guiding field is
$\Bol=1.4$~mT. The trap can be realized, for example, with a wire of
cross-section $4~\mu\mbox{m}\times4~\mu$m. In this case, the required wire current is 1.6~A, and the trap-wire distance is $\sim 2~\mu$m, depending slightly on the shape of the cross-section.

Today's strongest traps have not yet approached this maximum practical value: gradients in chip traps are typically $\sim 250$~T/m. For sub-micrometer structures, atom-surface interactions \cite{Henkel,Fortagh02,Leanhardt03,Jones03,Casimir} may impose more severe limits than the current density does. Nevertheless, it appears realistic to achieve trapping in the Tonks-Girardeau regime, as discussed below.

\section{Tonks-Girardeau gases in chip traps}
\label{sec:TG}

\subsection{The one-dimensional regime}

A gas is {\em quasi-one-dimensional} when the average energy per particle is much less than the energy of the first excited state $\hbar \omega_\perp$ in the transverse trap directions. (From here onward we will drop ``quasi-'' and simply refer the regime is one-dimensional or `1D' .) The average energy per particle has contributions from kinetic energy, potential energy, and from the inter-particle interactions (of order $\mu$, the chemical potential). At finite temperature, particles are excited by residual thermal energy (of order $k_\mathrm{B} T$). Thus in general, we can write the criterion to be in the 1D regime
\begin{equation}
       \{ k_{\mathrm{B}} T, \mu \} \ll \hbar \omega_\perp
\end{equation}
In the $T=0$ limit, which we will consider from here on, an equivalent
condition is that the amplitude of transverse zero point oscillations
$l_\perp=\sqrt{\hbar/m\omega_\perp}$ be much smaller than the
(density) coherence length $l_c=\hbar/\sqrt{m\mu}$, where $m$ is
the mass of the atom.


The 1D interaction strength is given by\footnote{Here and in the rest of this section we assume
$l_\perp > C a/\sqrt{2}$, as discussed in more detail in \S\ref{sec:config}.}
\begin{equation}
\label{eq:g1d}
g=2\hbar^2a/ml_\perp^2 = 2 \hbar \omega_\perp a. 
\end{equation}
Since $g$ increases linearly with
$\omega_\perp$, the strongly interacting TG regime is
more easily accessible with stronger confinement. The length scale
associated with $g$ is the one-dimensional scattering length
\begin{equation}
a_{\mathrm 1D} = - 2 \hbar^2 /m g = - l_\perp^2/a.
\end{equation}
Note that although the one-dimensional scattering length maintains its meaning in the 
scattering amplitude (\textit{i.e.} $f \propto (1 + i k a_{\mathrm 1D})^{-1}$, where $k$ is the wave vector), it cannot be interpreted like the 3D scattering length $a$. In fact, $a_{\mathrm 1D}$ is negative for repulsive ($g>0$) interactions, and the interaction strength is inversely proportional to $a_{\mathrm 1D}$.

\subsection{Review of Tonks-Girardeau theory}
\label{sec:TGtheory}

A Tonks-Girardeau (TG) gas is a one-dimensional ensemble in a low-density, strongly interacting limit. 
Since interactive energy far exceeds kinetic energies, the particles cannot overcome the two-body interaction potential, and are thus ``impenetrable bosons.'' The beautiful simplicity of this regime is that there is a one to one mapping from this strongly interacting system of bosons, onto an ideal system of fermions in the same one-dimensional potential. \cite{TGearly,Olshanii98} In the following paragraphs, we will present the criteria for being in the TG regime, and some properties of TG ensembles.

A weakly-interacting 1D gas has been achieved in several experiments \cite{quasi1Dexp}. Reaching the TG regime poses additional constraints on the {\it longitudinal} energy scales, but we will see that this requires an even stronger transverse confinement -- with oscillation frequencies approaching 1~MHz.

In a strongly interacting system, the interaction energy is much larger than the free-particle energy. For our case, two dimensionless parameters can indicate if this condition is fulfilled \cite{Petrov00}:
\begin{eqnarray}
\alpha & = & \frac{\ell_{\rm z}}{|a_{\mathrm 1D}|} = m g \ell_z / 2 \hbar^2, \quad \mbox{and} \label{eq:alpha} \\
\gamma & = & 1/(nl_c)^2 = mg/\hbar^2 n, \label{eq:gamma}
\end{eqnarray}
where $\ell_{\rm z}=\sqrt{\hbar/m \omega_{\rm z}}$ is the extent of the longitudinal
ground state, and $n$ is the number density (or local density, in the case of a non-uniform potential). 
The first
parameter, $\alpha$, relates to the ratio of the interaction energy
(characterized by $g$) to the potential energy (characterized by
$\omega$): if we write $\epsilon_{int} = \hbar^2/m a_{\mathrm 1D}^2$, then
$\alpha^2 = \epsilon_{int}/\hbar \omega_z$. The second parameter,
$\gamma$, is the ratio of the chemical potential $\mu$ to the kinetic
energy $\epsilon_{kin} \approx \hbar^2 n^2/m$. This form of
$\epsilon_{kin}$ assumes that particles fill the trap in a fermionic way,
such that the $N^{th}$ particle has $N$ nodes in its wave function and
thus a wave number of $N/L$, where $L$ is the length of the uniform
potential. For a harmonic potential, Eq.~(\ref{eq:gamma}) is replaced by
\cite{Dunjko}
\begin{equation}
\eta^{-1} = \frac{1}{n^{0} \left|a_{\rm 1D}\right|} = 
\left( \frac{8 m \omega_\perp^2 a^2}{3 \hbar \omega_{\rm z} N} \right)^{2/3},
\end{equation}
where $n^{0}$ is the peak density in the 1D Thomas Fermi regime. 

The properties of a gas in the TG regime have been discussed in many of the works cited above. We will cite two results here useful for the discussion in later sections. First, the longitudinal extent of the TG gas in a harmonic trap is \cite{MoreTonksTheory,Girardeau01,Dunjko}
\begin{equation}
R_{\rm TG} = \ell_{\rm z} [2\, N]^{1/2} .
\end{equation}
Second, at close range, the second-order correlation function decreases to \cite{Shlyapnikov03}
\begin{equation}
  \label{eq:g2large}
   g_2 (0)/n^2= 4\pi^2/3\gamma^2.
\end{equation}

Using the mapping theorem, one can use (the absolute value of) the ground state wavefuction of an ideal Fermi gas to find the distance at which this decrease occurs \cite{TGg2,Drummond}. For a uniform potential, the anti-bunching length scale is $L/N$, the average interparticle spacing \cite{Girardeau01}; for a harmonic trap, however, we find that anti-bunching occurs on a length scale $\sim \ell_{\rm z}/N$, which is $\sqrt{N}$ {\it smaller} than the average inter-particle spacing. Thus measuring the length scale of the dip in $g_2$ may be easier in a box-like potential than with a harmonic longitudinal confinement, as is discussed further in \S\ref{subsec:resolution}.

\subsection{Transverse confinement}
\label{sec:config}

Even though strong confinement helps us enter the TG regime, the confinement must not be so strong that $l_\perp \leq C a/\sqrt{2}$, where $C\approx 1.46$ \cite{Olshanii98}, else the scattering length will change sign. For $^{87}$Rb, this means that $\omega_\perp < 2 \pi \times 3.9$~MHz. Increase $\omega_\perp$ further if $a$ were reduced by a Feshbach resonance, however, let us choose for a target oscillation frequency $\omega_\perp = 2 \pi \times 1.1$~MHz, since this also constitutes a reasonable limit for the transverse oscillation frequency achievable in a chip trap at room temperature, as discussed in \S\ref{sec:maxFreq}. In the rest of this section, we will assume a wire of width $w=4~\mu$m and zero height\footnote{For a  real wire with $4~\mu$m height, the results would be slightly more  benign in that the trap center would be placed farther away from the  surface.}, so that we can use the analytical formulas of \S\ref{sec:broad}, and assume $I=1.6$~A. For this wire, the required $B'$ occurs at $x_0=2.5~\mu$m (Eq.~\ref{eq:gradBBroad}). An external homogeneous field $\Bot=108.4$~mT will place the trap center at this $x_0$ (Eq.~\ref{eq:BBroad}).

\subsection{Longitudinal confinement}

\begin{figure}[tb!]
  \begin{center}
    \includegraphics[width=0.8\columnwidth]{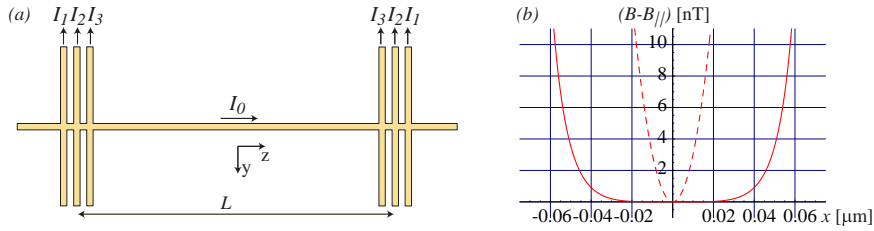}
  \end{center}
  \caption{Elongated Ioffe-Pritchard trap for the Tonks-Girardeau
    regime. $(a)$ Wire layout. $I_0$, together with the external field
    $\Bot$, provides strong transverse confinement. The other
    currents provide longitudinal confinement. $(b)$ Trapping potential
    (magnetic field modulus) provided by this configuration with the
    following parameters: width of all wires and of gaps between
    wires: $w=4~\mu$m; length $L=200~\mu$m; $\vec{\Bol}=\vecx\times
    1.4~$mT, $\vec{\Bot}=\vecz\times 108.4$~mT,
    $I_0=1.6$~A. The longitudinal confinement can be chosen to be
    harmonic or more box-like by appropriate choice of $I_1\ldots
    I_3$. For the dashed line, $I_2=1$~mA and $I_1=I_3=0$, resulting
    in harmonic confinement with a frequency of 10~Hz for the
    $F=2,m=2$ state of $^{87}$Rb. The solid line results when
    $I_1=0.6$~mA, $I_2=-1$~mA, $I_3=0.4$~mA. This configuration
    cancels the quadratic part of the longitudinal potential in the
    center of the trap.}
  \label{fig:longTrap}
\end{figure}
The parameters discussed in \S\ref{sec:config} only concerned the transverse confinement. To create a trapped 1D gas, some weak confinement must be added in the $\vecz$ direction. This can be achieved easily by adding two ``pinch'' currents along $\vecy$ at $z=\pm L/2$. This ``H''-shaped configuration \cite{Reichel00} is a generalization of the ``Z''-shaped wire trap, which was first described in \cite{Reichel99} and is widely used in the chip trap community. The magnetic fields of the ``pinch'' wires have $x$ and $z$ components only. Close to the trap minimum, the dominant contribution is along $\vecz$, (the direction of the guiding field $\Bol$), and increases as one moves towards $z=\pm L/2$.

This method of creating longitudinal confinement is not limited to two conductors. More conductor pairs can be added, in which case it becomes possible to control the shape of the longitudinal confinement. We thus arrive at the configuration shown in Fig.~\ref{fig:longTrap}$(a)$. The distance $L$ and the currents
\footnote{Note that it is possible to adjust the currents  independently in spite of the conductor crossings, provided that  floating current sources are used. In the limit of thin conductors,  the resulting current distribution is the same as that of isolated  conductors. However, multilayer chips have also been used for this  purpose \cite{Long03}, and are required if two currents cross more  than once.}
in the ``pinch'' conductors can be varied to obtain the desired longitudinal confinement, as shown in Fig.~\ref{fig:longTrap}$(b)$. Note also that with this configuration, longitudinal and transverse confinement can be adjusted independently.

\subsection{Possible Tonks-Girardeau parameters in a chip trap}
\label{sec:params}

Having shown we can create a strong transverse confinement and a wide variety of longitudinal confinements, we can now calculate the number of atoms $N^{\rm TG}$ for which we cross into the Tonks regime.

A harmonic trap with $\omega_z = 2 \pi \times 5$~Hz gives $\eta^{-1} = (N/N^{\rm TG})^{2/3}$, where $N^{\rm TG} = 1.3 \times 10^5$. For $N=30$, $\eta^{-1} = 265$, the spatial extent is $R^{\rm TG} = 37~\mu$m, and the inter-particle spacing is $n^{-1} = 2.0~\mu$m at the center of the trap. The chemical potential is $\mu/h = 150$~Hz.
 
For a box potential (or ``uniform'' potential) with $L=100~\mu$m and $\omega_\perp = 2 \pi \times 1$~MHz, $\gamma = N^{\rm TG}/N$, where $N^{\rm TG} = 9.2 \times 10^3$. For $N=30$, $\gamma > 300$, the average inter-particle spacing is $n^{-1} = 3.3~\mu$m, and the chemical potential is $\sim$50~Hz.

In conclusion, we see that for atom numbers $N \ll 10^3$, we can be deeply in the TG regime using a chip trap.

\section{Detection by discretization}
\label{sec:detection}

As shown above, the Tonks-Girardeau regime and realistic trap parameters demands a very small total number of atoms in the elongated trap. If the atomic distribution is to be measured with sufficient spatial resolution in order to determine its correlation function, the imaging system must have a detectivity approaching one atom per pixel, combined with a spatial resolution near the diffraction limit, over the whole longitudinal extent of the trap. No existing imaging systems fulfills all these requirements at once. Systems with single-atom detectivity per pixel, \textit{e.g.}~\cite{Schlosser01,Kuhr01}, are usually designed to collect fluorescence light with high numerical aperture optics. Consequently, they have a very small field of view and an insufficient number of pixels for our requirements. As a solution we propose to combine such a detector with the discretization and transport method described below, as shown in Fig.~\ref{fig:Conveyor}. The atomic distribution is broken up into a 1D chain of ``buckets''. As each bucket of trapped atoms passes in front of the detector, the number of atoms is counted, and a position distribution is built up. 

\subsection{Modulating the longitudinal confinement: Discretization and the conveyor belt} 

\begin{figure}[tb!]
 \begin{center}
  \includegraphics[width=\columnwidth]{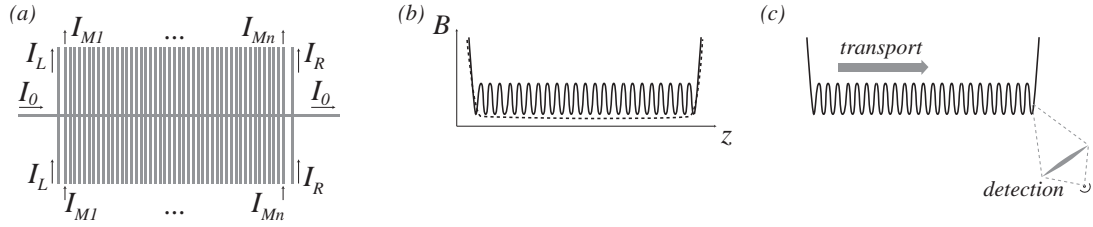}
 \end{center}
 \caption{Detection scheme for the TG gas. Motor currents 
  $I_{M1}\ldots I_{Mn}$, shown in $(a)$, are used to discretize the
  initial longitudinal potential (dashed line in $(b)$) into a chain
  of ``buckets'' (solid line). By modulating the currents, the
  buckets are made to slide past a single-atom detector $(c)$, where
  the number of atoms in each bucket is counted.}
 \label{fig:Conveyor}
\end{figure}
The same method that we have used above to create longitudinal
confinement also allows creation of a linear chain (a 1D lattice) of
potential wells (Fig.~\ref{fig:Conveyor}$(a)$ and $(b)$). In
particular, if an atomic gas is initially held in the elongated
potential of Fig.~\ref{fig:longTrap}$(b)$, the atomic distribution can
be discretized into a linear chain of buckets by switching on
additional currents. If the switching is done sufficiently fast
(\textit{i.e.}, fast enough to avoid tunneling, but not so fast as to
excite higher longitudinal lattice bands), the initial density
distribution will be frozen: the number of atoms in each bucket
reflects the local atomic density (averaged over the extent of the
bucket) that was present in the elongated trap before the
discretization.

With an appropriate time-dependent modulation of the $\vecy$ currents, the chain of minima of Fig.~\ref{fig:Conveyor}$(b)$ can be continuously moved along $\vecz$, as shown schematically in part $(c)$ of the figure. Obviously, the finer the discretization and the longer the transport distance, the larger the number of $\vecy$ currents must be. However, it is not necessary to control all these currents individually: They can be connected in groups.
This is the basic idea of the ``atomic conveyor belt'', which is described in \cite{Haensel01} and \cite{Reichel00}. In those implementations, instead of the many straight wires along $\vecy$, a pair of counter-wound wires were used to avoid multiple crossings with the long central wire. Although this makes the chip easier to produce, the resulting potential is slightly more complicated. Recently however, the use of a multilayer chip was demonstrated in Munich to realize exactly the fundamental transport scheme of Fig.~\ref{fig:Conveyor} \cite{Long03}. This experiment will be reported in more detail in a future publication. In the present context, the important point is that the scheme of Fig.~\ref{fig:Conveyor} can indeed be used to discretize an elongated potential into multiple wells, and that these wells can subsequently be transported in a controlled way. We will use this scheme to discretize a 1D atomic cloud and transport the resulting chunks to a single-atom detector, in order to achieve spatially resolved detection with just a single atom-counting detector.

We have shown above how this discretization and transport mechanism can be implemented in a magnetic chip trap. Note, however, that the method itself can also be applied to optical traps. In that case, the TG gas would initially be prepared in a dipole trap (optical wavelength $\lambda$). By suddenly switching to a standing-wave configuration, the atomic distribution is discretized with a resolution of $\lambda/2$; a controlled detuning between the two counter-propagating beams transports the atoms to a fluorescence detector. Exactly this way of transporting and counting individual atoms has already been demonstrated with thermal atoms\cite{Kuhr01}. Of course, combining the two parts still represents a daunting task, but the experimental feasibility seems reasonable. 

\subsection{Spatial resolution and detector requirements}
\label{subsec:resolution}
In this scheme, the detector itself no longer needs to have any
spatial resolution (the CCD can be replaced by a single photodiode).
Instead, the spatial resolution is determined by the spatial period of
the discretization. 

Before we discuss the limits of this discretization, let us briefly
consider the required resolution of the detector {\em optics}. This
resolution must be high enough so that only one bucket is imaged onto
the photodiode. In the case of an optical standing wave, the bucket
size is the period of the standing wave, \textit{i.e.} $\lambda/2$ -- a
resolution that is hard to achieve if $\lambda$ is of the same order
as the wavelength of the fluorescence light. In the case of the chip
trap, the period of the modulation wires does not need to be constant:
It can be small in the trapping region, but larger near the
detector. The bucket chain is then stretched while it is being
transported, and detector optics with poor resolution can be used.

We now come back to the resolution limit of the discretization. If an
optical potential is used, this resolution is $\xi=\lambda/2$. Whether
the same resolution can be achieved for a chip trap is still an open
question. In order to achieve a resolution $\xi$, the first
requirement is to fabricate conductors with a width and spacing of
$\xi/2$ or better. Although the minimum conductor width and spacing
for existing chip traps is $\sim 1~\mu$m \cite{FolmanReview02},
photolithography is known to work well at much smaller scales -- the
standard in commercial microchip manufacturing is currently moving
from 130~nm to 90~nm. Thus, chip fabrication will not be the
main obstacle. However, a second condition is that the trap-surface
distance must also be of order $\xi$ -- at larger distance, the
periodic structure in the potential would be averaged out. Recent
measurements of trap lifetime near surfaces \cite{Jones03,Harber03}
indicate that 100~ms lifetimes will still be possible at 1~$\mu$m
from a thick copper surface, and still longer lifetimes for thinner
metalization layers and for metals with higher resistivity. At still
smaller distances, the attractive Casimir-Polder potential must be
taken into account. \cite{Casimir} Thus, although it is too early to predict how far
the chip trap resolution can ultimately be pushed, it seems reasonable
to expect a resolution of $2~\mu$m.

Considering the result of \S\ref{sec:TGtheory} for a {\em uniform trap},
this resolution would be adequate to resolve the
inter-particle repulsion characteristic of the TG regime, for roughly
$N=30$. For a {\em harmonic trap}, however, with the example parameters
given in \S\ref{sec:params}, the drop in $g^{(2)}$ would have a width
of less than $0.2~\mu$m, smaller than the resolution $\xi$ given
above. Thus observing the dip in the two-body correlation function
would require an even smaller $N$ and weaker axial confinement: at
$\omega_z = 0.25$~Hz and $N=10$, $l_z/N$ is roughly 2~$\mu$m. These
parameters are unrealistically low, pointing out an advantage of
uniform traps. By contrast, even with the harmonic trap parameters in
\S\ref{sec:params} and the proposed detection method, one could
carefully measure the density distribution characteristic of
the TG regime.

\section{Discussion}
\label{sec:conclusion}

The trap and detection mechanism proposed in the above sections have
not been realized, but provide a vision of interesting physics that
motivate the continued improvement of microfabricated traps and
transport devices for neutral atoms. Not only are excellent trap
parameters possible for the TG regime, but a well-suited detection
mechanism might only be possible with an integrated atom chip.
Detection is certainly the most challenging part of realizing a Tonks
gas on an atom chip. Corrugations of the potential
\cite{Fortagh02,Leanhardt03,Jones03} would lead to limitations if the chip was
produced with current atom chip fabrication methods. However, a recent
investigation \cite{Aspect04} of these corrugations suggests that this
problem can be mitigated by using lithography processes with higher
resolution.

The method of ``freezing'' the distribution with a resolution comparable to the inter-particle spacing would allow the measurement not only of the density distribution, but also of the particle-particle correlation functions. The inter-particle repulsion in this strongly interacting regime is directly visible in such correlation functions.

Finally, addressing several practical issues is essential to realizing the TG regime. For instance, the path from a normal Bose condensate to the TG regime may be problematic, due to excitations or particle loss. Also, surface-atom interactions are critical to understand and control in this high-confinement regime, where atoms are within microns of the surface. The successful integration of all the chip technologies discussed is an ongoing project in Munich.

\begin{acknowledgements}
J.R.\ thanks Philippe Grangier, Christophe Salomon and Maxim Olshanii
for stimulating discussions at Les Houches, which inspired the
detection scheme presented here, and Wolfgang H\"ansel for
help with the magnetic field simulations. J.T.\ thanks Maxim Olshanii for discussions concerning the correlation length of a TG gas. The authors also thank H\'{e}l\`{e}ne Perrin and Bruno Labruth-Tolra for a critical read of the manuscript.
\end{acknowledgements}


\end{document}